\documentclass[aps,prl,showpacs,twocolumn,superscriptaddress,amsmath]{revtex4}
\usepackage{graphicx}
\usepackage{amssymb}
\usepackage{bm}
\begin{document}
\setcounter{page}{1}

\title{Finite-temperature phase transitions in a two-dimensional
boson Hubbard model}
\author{Min-Chul Cha}
\affiliation{Department of Applied Physics, Hanyang University, Ansan 426-791, 
Korea}
\author{Ji-Woo Lee}
\affiliation{Department of Physics, Myongji University, Yongin 449-728,
Korea}

\begin{abstract}
We study finite-temperature phase transitions in a two-dimensional
boson Hubbard model with zero-point quantum fluctuations
via Monte Carlo simulations of quantum rotor model,
and construct the corresponding phase diagram.
Compressibility shows a thermally activated gapped behavior 
in the insulating regime.
Finite-size scaling of the superfluid stiffness clearly shows
the nature of the Kosterlitz-Thouless transition.
The transition temperature, $T_c$, confirms a scaling relation 
$T_c \propto \rho_0^x$ with $x=1.0$.
Some evidences of anomalous quantum behavior at low temperatures are presented.

\end{abstract}



\maketitle

Recently quantum phase transitions\cite{Sachdev,Sondhi97}
have drawn a lot of attention in systems of interacting particles.
Typically strong interactions suppress the itineracy of particles
to induce a strongly correlated insulating phase,
whereas with weak interactions a conducting phase is stable.
The criticality of these zero-temperature phase transitions
can be investigated at low, but finite, temperatures.
How quantum fluctuations associated with 
a quantum critical point(QCP) have influence on
phases at finite temperatures
\cite{Coleman05,Chakravarty05,Sachdev97}
is a theoretically interesting and an experimentally relevant question.

At finite temperatures, it is expected that
a quantum phase transition turns into a classical one
with the same order parameter or disappears.
Remnant quantum fluctuations near a QCP
may bring anomalous properties\cite{Coleman05},
which can be captured by scaling relations,
and lead to crossover behaviors as temperature rises.
Some possibilities such as reentrant behaviors
due to the interplay of quantum and thermal fluctuations
have been proposed\cite{Kim90}.

These issues can be clarified by direct investigations of a generic quantum
mechanical model.
So far most of the theoretical investigations heavily rely on the exact solution of 
the quantum Ising model, available strictly in one dimension\cite{Chakravarty05}.
Interacting bosonic systems simulated via Monte Carlo methods, not suffering
from negative sign problems, will be an ideal place to study these problems.
In previous works, a quantum $XY$ model,
equivalent to hard-core bosons at half-filling, 
showed the Kosterlitz-Thouless(KT) transition\cite{Kosterlitz73}
at finite temperature in two dimensions\cite{Ding90,Harada98}.
In the model with nearest neighbor repulsion, destruction of the solid order
as well as the superfluidity by thermal fluctuations was observed
\cite{Schmid02}.
However, generic finite-temperature phase diagrams have not been constructed.

\begin{figure}[b]
\includegraphics[width=8.0cm,height=6.7cm]{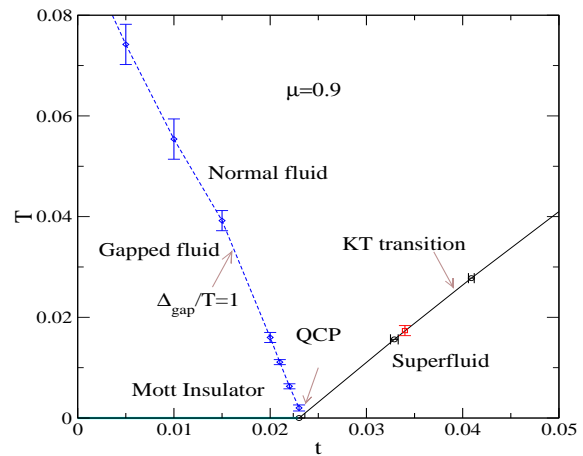}
\caption{(Color online)
Phase diagram on the space of
hopping strength, $t(=\sqrt{n_0(n_0+1)}w)$, and temperature, $T$, in unit of $U$.
The solid line denotes the classical phase transitions,
which terminates at a QCP at $T=0$.
The dotted line represents crossover between gapped fluid and
normal fluid.
}
\label{fig:phase_diagram}
\end{figure}

In this work, we investigate the thermally driven phase transitions of
a two-dimensional quantum rotor model,
which is believed to share the same critical properties
of a soft-core generic boson Hubbard model\cite{Fisher89},
via Monte Carlo simulations.
The results are summarized in the phase diagram
as shown in Fig.~\ref{fig:phase_diagram}.
Finite-size scaling properties of the superfluid stiffness
confirm that the nature of the classical phase transition
associated with the destruction of superfluidity is
consistent with that of the KT transition, and clearly
support the scenario of the universal jump at the critical point\cite{Nelson77}.
Finite temperature, $T$, sets the size in the temporal direction,
leading to a scaling behavior\cite{Chakravarty05,Fisher89}
$T_c \propto \rho_0{}^x$ with $x=1.0$,
where $T_c$ is the transition temperature and $\rho_0$ is the superfluid
stiffness at zero temperature.
The compressibility diverges at the transition.
In the insulating regime at low temperature,
thermally activated behavior of the compressibility with a finite energy gap
is observed.
Some anomalous dependence of energy and specific heat on $T$,
possibly due to quantum fluctuations, are observed for $T < 0.25 U$.

The Hamiltonian of a boson Hubbard model reads
\begin{eqnarray}
H={U \over 2} \sum_j  n_j(n_j-1) -\mu \sum_j n_j
- w \sum_{<ij>} (b_i^\dagger b_j + b_j^\dagger b_i ),
\label{eq:bH}
\end{eqnarray}
where $b_j (b_j^\dagger)$ is the boson annihilation(creation) operator
at the $j$-th site, and $n_j$ is the number operator.
$U$ and $w$ stand for the strengths of the on-site repulsion and
of the nearest neighbor hopping, respectively,
and $\mu$ is the chemical potential.

It is convenient to put $\mu/U+1/2 = n_0 + {\bar n}$ 
with an integer $n_0$ and $-1/2 < {\bar n} \le 1/2$
so that $n_0$ represents the background number of bosons per site
and $\bar n$ is a charge offset.
When ${\bar n}=0$, the density of bosons is fixed to a commensurate filling
across the transition.
For non-integer ${\bar n}$, however, an integer filling in a Mott insulator
shifts to a non-integer filling in a compressible fluid.
We study the phase transition of the latter case in (2+1)-dimensional
$L\times L\times L_\tau$ square lattices,
where $L$ denotes the size in a spatial dimension
and $L_\tau$ in the temporal dimension.

Since the phase transition of the model in Eq.~(\ref{eq:bH})
is characterized by the establishment of phase coherence,
we may rewrite the Hamiltonian 
in terms of the phase angle $\theta_j$ of bosons by 
replacing $b_j (b_j^\dagger) = \sqrt{n_j} e^{-i \theta_j} 
(\sqrt{n_j+1}e^{i\theta_j})$
with $n_j = {1 \over i}{\partial \over \partial \theta_j}$.
Under the assumption that the nature of the transition is governed only
by the fluctuations of $\theta_j$, not those of the hopping strength,
we replace $n_j \to n_0$ so that $b_j (b_j^\dagger) = \sqrt{n_0} e^{-i \theta_j} 
(\sqrt{n_0+1}e^{i\theta_j})$.
Then, the Hamiltonian is reduced to a quantum rotor model
\begin{eqnarray}
H={U \over 2} \sum_j  n_j(n_j-1) -\mu \sum_j n_j
- 2t \sum_{<ij>} \cos(\theta_i - \theta_j),
\end{eqnarray}
where $t=\sqrt{n_0(n_0+1)} w$.
Here we take the number of bosons $n_j \geq 0$.

Through a path integral mapping, we can construct the corresponding 
classical action\cite{Wallin94}
\begin{eqnarray}
S=\sum_r\frac{\epsilon U}{2} J_r^\tau(J_r^\tau-1)
-\epsilon\mu J_r^\tau
-\ln{I_{J_r^x}}(2\epsilon t)-\ln{I_{J_r^y}}(2\epsilon t)
\label{eq:action}
\end{eqnarray}
with the partition function
\begin{eqnarray}
Z=\sum_{\{\vec J_r\}}^{\nabla\cdot\vec J=0} e^{-S[\vec J]},
\label{partition_function}
\end{eqnarray}
where $\epsilon=\beta/L_\tau$ is a lattice constant in the imaginary
time axis for an inverse temperature $\beta$,
$\vec J_r$ is an integer current at site $r=(j,\tau)$
with a spatial index $j$ and a temporal index $\tau$,
which is conserved at each site as denoted by $\nabla\cdot\vec J=0$,
and $I_m(x)$ is the modified Bessel function given by the relation
$e^{K\cos\theta}=\sum_{m= -\infty}^{\infty}I_m(K)e^{im\theta}$.
In this work, we investigate the properties of the model in Eq.~\ref{eq:action}
via Monte Carlo simulations using a recently proposed worm algorithm
\cite{Alet03}.
In order to reduce the systematic errors in discretizing
the imaginary time axis, we need to take $\epsilon \sqrt{tU}\ll 1$.
We take $U\epsilon =$ 0.5 - 2 for $t\ll U$
and set the energy unit $U=1$.

The superfluid stiffness in a finite system is given by\cite{Wallin94} 
\begin{eqnarray}
\rho_L=\beta^{-1}L^{2-d} \langle W_x^2 \rangle,
\end{eqnarray}
where $W_x=L^{-1}\sum_r J_r^x$ and $\langle ... \rangle$ denotes
the averages over the probabilites determined 
by the partition function of Eq.~(\ref{partition_function}),
and $d$ is the spatial dimensionality.
Similarly the compressibility is
\begin{eqnarray}
\kappa=\beta L^{-d}[\langle N^2\rangle-\langle N\rangle^2],
\end{eqnarray}
with $N=L_\tau^{-1}\sum_r J_r^\tau$.
The energy expectation is given by
\begin{eqnarray} 
\langle H\rangle
= L_\tau^{-1}\langle \frac{\partial S}{\partial \epsilon}\rangle\ ,
\end{eqnarray}
and the specific heat is $C_V=L^{-d}(\partial\langle H\rangle/\partial T)$.


\begin{figure}
\includegraphics*[width=8.0cm, height=6.7cm]{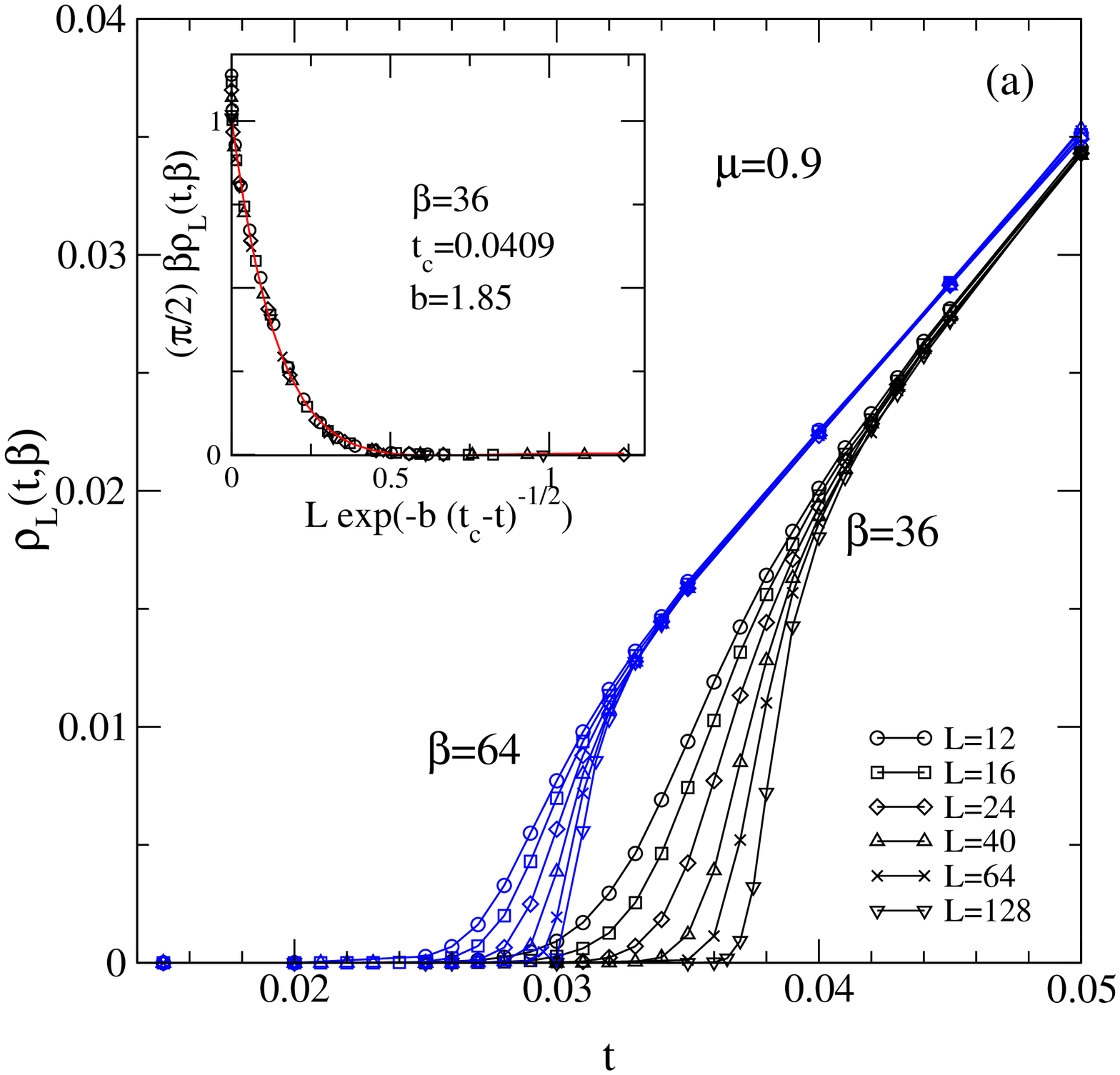}
\includegraphics*[width=8.0cm, height=6.7cm]{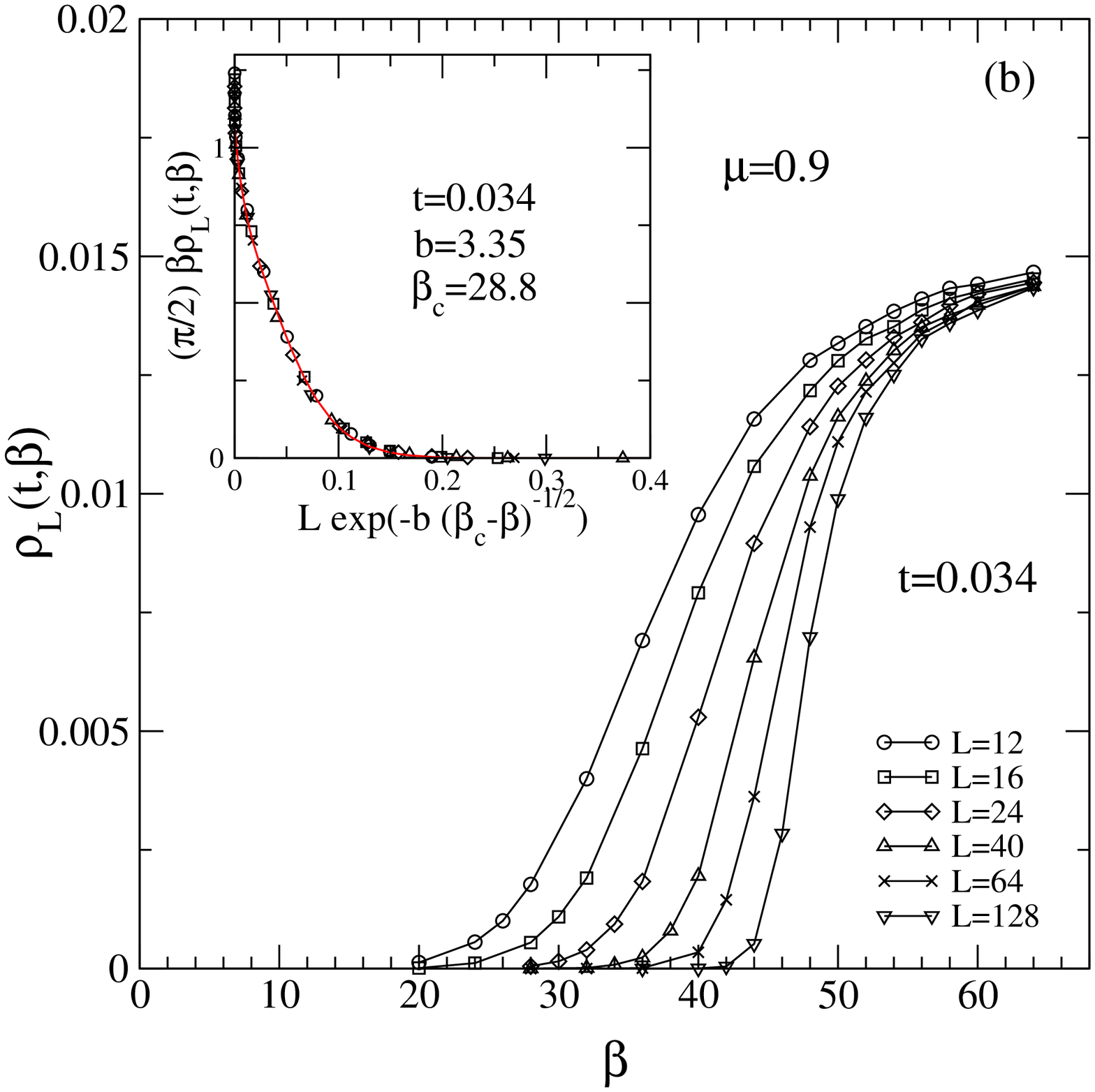}
\caption{(Color online)
Finite-size scaling behaviors of the superfluid stiffness
as a function of (a) hopping strength and (b) temperature.
For both cases, data collapsing onto a single curve
works fine in terms of the scaling parameter $L/\xi$ as shown in insets,
consistent with the nature of
the KT transition and the universal jump at the critical point.
}
\label{fig:rho_mu09}
\end{figure}

We consider the case for $\mu=0.9$ so that $n_0=1$ and $\bar n = 0.4$.
Figure~\ref{fig:rho_mu09} shows the finite-size scaling behavior of the superfluid
stiffness as a function of (a) $t$ and (b) $\beta$.
Finite-size scaling properties of the transition can be obtained by plotting
the curves in terms of a scaling variable $L/\xi$, where $\xi$ is the
correlation length.
Here we assume an essential singularity\cite{Kosterlitz74}
$\xi \sim \exp(b \delta^{-1/2})$,
where $\delta=t-t_c$ (or $\beta-\beta_c$)
is a tuning parameter and $b$ is a non-universal scaling factor.
In terms of this scaling variable, we obtain high-quality
data collapsing onto a single curve for different sizes,
consistent with the nature of the KT transition.
The scaling behavior also supports the scenario of the universal
jump of the superfluid stiffness\cite{Nelson77}, $(\pi/2)\beta_c\rho_\infty=1$,
at the critical point in the thermodynamic limit.
By extrapolating the single curves to the critical point,
we find that 
$(\pi/2)\beta_c\rho_\infty\approx$ (a)$1.01$ and (b)1.06. 
These numbers are, however, 
sensitive to fitting parameters $b$ and $t_c (\beta_c)$.

\begin{figure}
\includegraphics*[width=8.0cm, height=6.7cm]{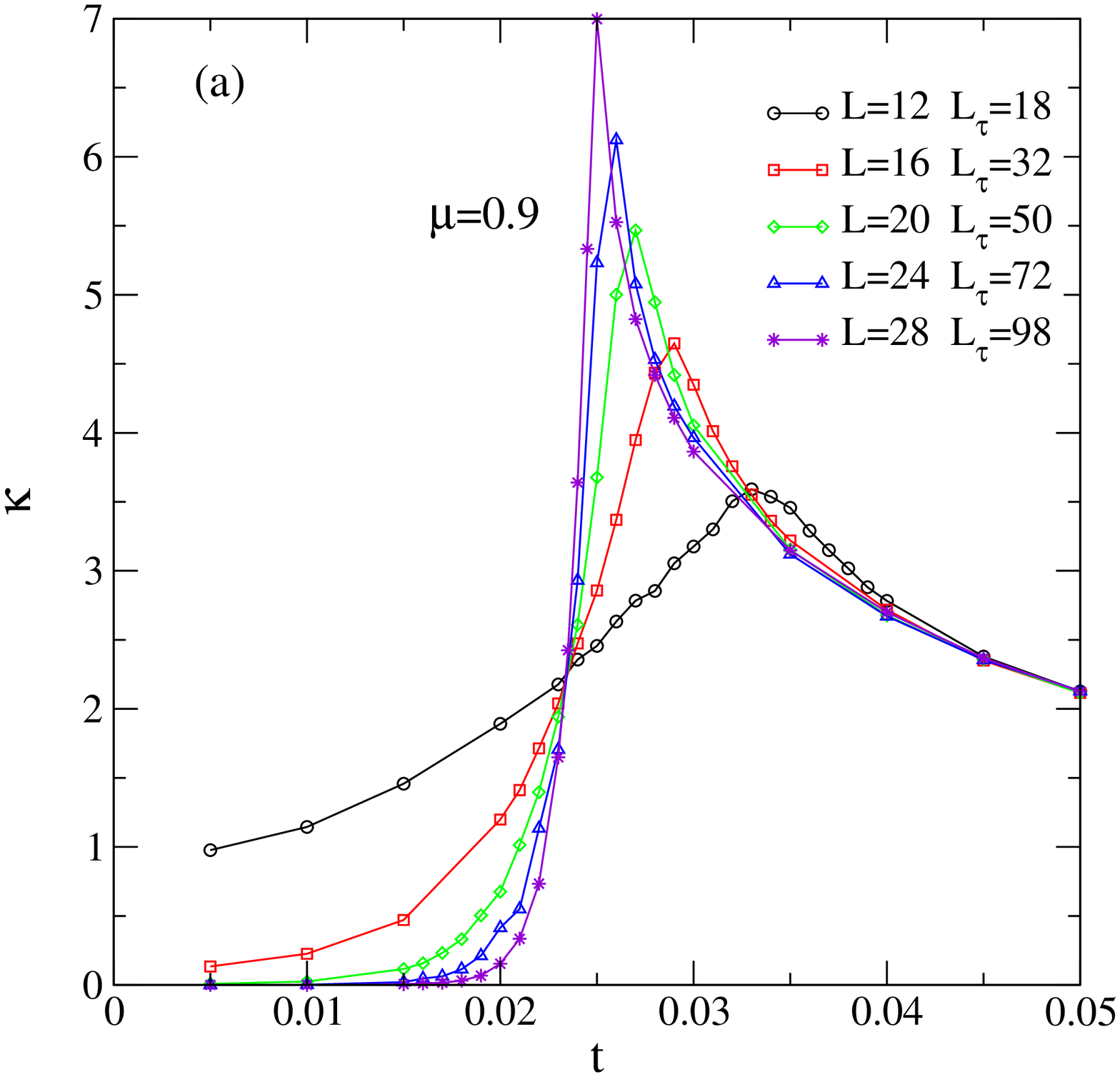}
\includegraphics*[width=8.0cm,height=6.7cm]{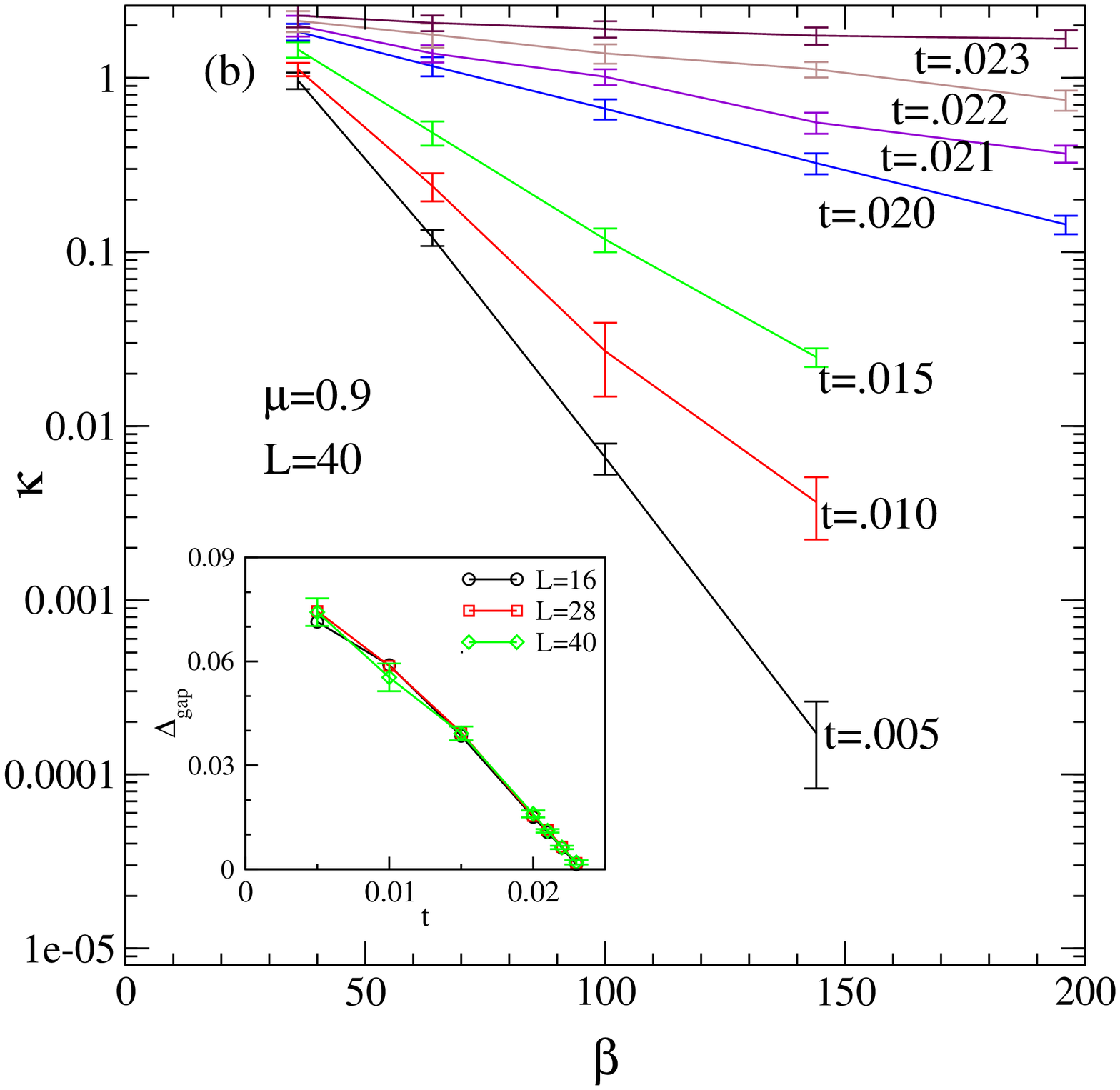}
\caption{(Color online)
(a) Compressibility of the boson Hubbard model
shows behavior of the GSIT with $z=2.0$, diverging at the transition.
(b) In the insulating regime, we have thermally activated behaviors,
$\kappa \sim e^{-\Delta_{\rm gap}/T}$, from which $\Delta_{\rm gap}$ 
can be evaluated.
Inset: $\Delta_{\rm gap}$ as a function of $t$, vanishing at the QCP.
}
\label{fig:compressibility}
\end{figure}

Figure~\ref{fig:compressibility}a shows the behavior of the compressibility.
The finite-size scaling ansatz of the compressibility is written in the form
\begin{eqnarray}
\kappa =L^{z-d}{\tilde X}_\kappa (L(t-t_c)^{1/\nu}, \beta/L^z),
\end{eqnarray}
where ${\tilde X}_\kappa$ is a dimensionless
scaling function and $z$ is the dynamical critical exponent.
For the generic superfluid-insulator transition(GSIT), $z=2$
is expected\cite{Fisher89}.
The crossing behavior of the compressibility curves for different sizes
at the critical point $t_c^0=0.023\pm 0.001$,
therefore, represents the scaling properties near the QCP,
where $t_c^0$ is the critical hopping strength at zero temperature.
For different values of $\mu$, we have similar results with $t_c^0$ just shifted.

We find that the compressibility diverges at the transition. 
In the superfluid side, $\kappa \sim 1/(t-t_c^0)$.
This strongly supports that the longe-range density fluctuations
drive the transition.
In the insulating side, the compressibility has an activated form
$e^{-\Delta_{\rm gap}/T}$ with a finite energy gap $\Delta_{\rm gap}$.
This dependence is shown in Fig.~\ref{fig:compressibility}b
for different $t$, from which
we can calculate $\Delta_{\rm gap}$ as shown in the inset.
For small $t$, we need a large number of Monte Carlo steps to obtain
equilibrium and have bigger error bars in determination of $\Delta_{\rm gap}$. 
The gap vanishes around $t = t_c^0$ as expected.

Thus we have a so-called 'V-shaped' phase diagram (Fig.\ref{fig:phase_diagram}).
In the insulating regime, the Mott insulator exists at $T=0$,
which turns into an activated gapped fluid with a finite
energy gap at low temperature.
It gradually disappears in a high-temperature normal fluid.
This crossover line can be specified by the condition 
$\Delta_{\rm gap}/T \approx 1$.
The phase coherence in a superfluid
at $T=0$ is destroyed by quantum fluctuations
to form a QCP or by thermal fluctuations at $T>0$
to define classical phase transitions.
The phase boundary in Fig.~\ref{fig:phase_diagram} is obtained by
tuning $t$ for given $T$ (black circles)
as well as by tuning $\beta$ for a given $t$ (red squares).
Note that the phase boundary
follows a scaling relation $T_c \propto |t-t_c^0|^{z\nu}$,
which implies that $\beta$ determines the correlation length
in the temporal direction,
where $\nu$ is the correlation length critical exponent.
The boundary in Fig.~\ref{fig:phase_diagram} is consistent with the expectation
$z\nu=1$\cite{Fisher89} for the GSIT.

\begin{figure}[b]
\includegraphics[width=8.0cm,height=6.7cm]{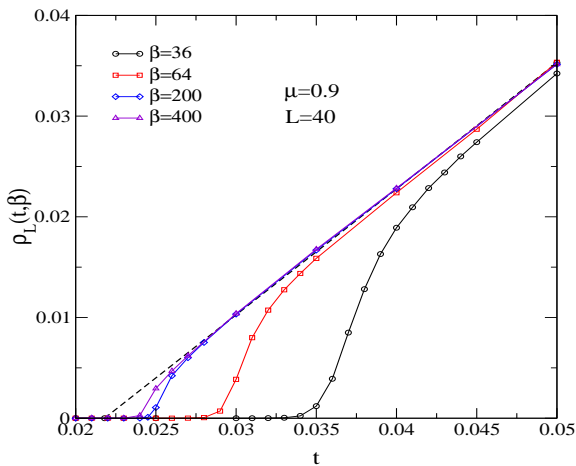}
\caption{(Color online)
Superfluid stiffness for different $\beta$.
As $\beta$ increases, the size dependence becomes smaller.
This allows us to extrapolate the curves to obtain zero-temperature
superfluid stiffness, $\rho_0$, in the thermodynamic limit
as denoted by dotted line.
It shows that $\rho_0 \propto |t-t_c^0|$ with $t_c^0\approx 0.22$.
}
\label{fig:rho0}
\end{figure}

It is interesting to check the predicted scaling relation
\cite{Chakravarty05,Fisher89} $T_c \propto \rho_0{}^x$ in this model.
Figure~\ref{fig:rho0} shows that the zero-temperature superfluid stiffness
$\rho^0$, denoted by dotted line, which obtained via extrapolation of values at
$T >0$, follows $\rho^0 \propto |t-t_c^0|$, implying that $x=1.0$.
It is consistent with the hyperscaling argument\cite{Fisher89} suggesting
$x=z/(d+z-2)$.

\begin{figure}[b]
\includegraphics[width=8.0cm]{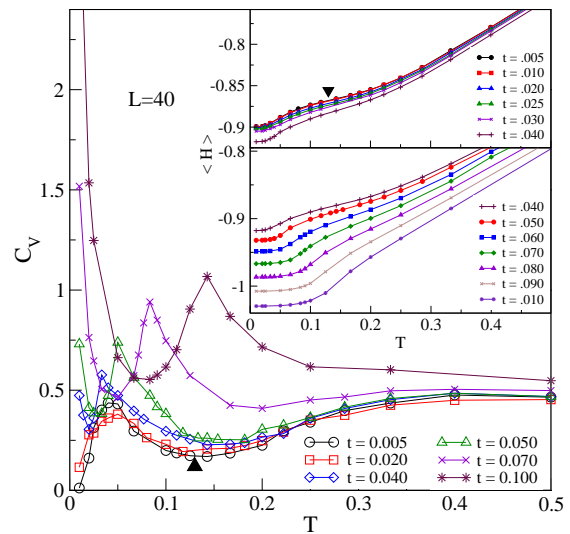}
\caption{(Color online)
Specific heat, $C_V$, as a function of $T$ for different $t$.
Sharp rises in the conducting regime, 
signature of the superfluid transition,
or round up-rise of $C_V$ in the insulating regime are followed
by indents which disappear in high temperature region, $T \gtrsim 0.25$.
Insets: 
The curves of the energy expectation values, $\langle H\rangle$, have bumps
at low temperatures possibly due to the effects of quantum fluctuations.
}
\label{fig:Cv}
\end{figure}

We expect that this quantum criticality
disappears as temperature rises,
which means quantum fluctuations possibly leave some tracks
in bulk properties at low temperatures.
Figure~\ref{fig:Cv} shows the specific heat, $C_V$,
and the energy expectation values, $\langle H\rangle$,
as a function of $T$ for different $t$.
Sharp rises of $C_V$ in the conducting regime 
or round up-rises in the insulating regime
are followed by indents, regions indicated by $\blacktriangle$,
which apparently represent anomalous behavior due to quantum fluctuations
and disappear at high temperatures for $T \gtrsim 0.25$.
This feature strongly suggests a crossover in normal fluid
from quantum mechanical to classical regime.
Similarly the curves of $\langle H\rangle$ show bumps, indicated by
$\blacktriangledown$, only in the range where
quantum critical fluctuations are expected to have effects.

In summary, we have investigated the phase transitions at finite temperature
in a two-dimensional quantum rotor model
in which intrinsic zero-point fluctuations are present.
Finite-size scaling of the superfluid stiffness shows
an essential singularity of the KT phase transition
and the universal jump at the critical point.
The compressibility diverges at the transition.
In the insulating regime, the compressibility shows a thermally activated behavior,
$\kappa \sim e^{-\Delta_{\rm gap}/T}$,
from which we can successfully evaluate the gap.
This indicates that the insulating behavior at low temperature
gradually crosses over to the behavior of normal fluid as temperature increases.
The transition temperature $T_c$ shows a scaling behavior $T_c \propto |t-t_c^0|$,
showing that finite $T$ limits the length of quantum fluctuations
in the temporal direction, and a hyperscaling relation $T_c \propto \rho_0$.
The behavior of the specific heat and the energy suggests that,
as temperature rises, quantum critical regime near a QCP crosses over
to classical regime.

MCC would like to thank Gerardo Ortiz for helpful discussions
and the hospitality of Department of Physics, Indiana University,
where parts of this work were carried out.
This work was supported by Korea Research Fund grant No. R05-2004-000-11004-0.

\end{document}